\documentclass[12pt]{iopart}
\usepackage{iopams} 
\usepackage[usenames,dvipsnames]{xcolor} 
\begin{document}

\title{Group theoretical formulation of free fall and projectile motion}

\author{Koray D\"{u}zta\c{s}}

\address{Physics Department, Eastern Mediterranean  University, Famagusta, North Cyprus via Mersin 10, Turkey}
\ead{koray.duztas@emu.edu.tr}
\vspace{10pt}

\begin{abstract}
In this work we formulate the group theoretical description of free fall and projectile motion. We show that the kinematic equations for constant acceleration form a one parameter group acting on a phase space. We define the group elements $\phi_t$ by their action on the points in the phase space. We also generalize this approach to projectile motion. We evaluate the group orbits regarding their relations to the physical orbits of  particles and unphysical solutions. We note that the group theoretical formulation does not apply to more general cases involving a time dependent acceleration.  This method improves our understanding of the constant acceleration problem with its global approach. It is especially beneficial for students who want to pursue a career in theoretical physics.
\end{abstract}

%
\vspace{2pc}
%
%
%
%

\section{Introduction}                                
\label{sec:level1}  
When does an object hit the ground if we release it from a certain height $h$? One of the greatest accomplishments in the history of science is Galileo's statement regarding this issue; i.e. the accelerations of freely falling objects are independent of their masses. This result is based on a thought experiment where one drops a heavy and a light object connected by a spring, from a height $h$. If heavier objects were to fall faster as assumed by Aristotle, the lighter object would fall slower and the string would pull up the heavier object to slow it down. On the other hand the system as a total is heavier than the heavy object, so it should fall faster. This contradiction is sufficient to invalidate Aristotle's assumption that the velocity of free fall is proportional to mass. Either by dropping weights from the leaning tower of Pisa, or by purely theoretical arguments, Galileo concluded that all bodies falling in vacuum experience exactly the same acceleration, and the distance fallen only depends on the square of the elapsed time~\cite{galileo}. This equivalence principle paved the way for Newton's formulation of gravity. Einstein's general theory of relativity which identifies gravity with the background geometry,  is actually based on the same equivalence principle. The validity of the equivalence principle has been verified up to very high precision. (see e.g. \cite{galileoexp})

Our ability to predict the ``future'' based on the knowledge of the ``present'' relies on the fact --or the assumption-- that the laws of nature are written in the mathematical language~\cite{wigner}. This principle allows us to describe the motion of free fall near the surface of the Earth as a constant acceleration problem. For that purpose we should also assume that the acceleration due to gravity does not considerably vary with the altitude at short distances, and the air resistance can be neglected. To the extent that these assumptions are justified, we can use the kinematic equations for constant acceleration which are directly derived from calculus.
\begin{eqnarray}
&& V=V_0 + at \label{fall1} \\
&& y_{\rm{fin}}= \frac{at^2}{2} + V_{\rm{in}}t + y_{\rm{in}} \label{fall2}
\end{eqnarray}
where $a$ denotes the constant acceleration, $V$ is velocity and $t$ is the time elapsed. We can combine the free fall problem with a  constant velocity in $x$ direction to get the projectile motion. A typical problem in the introductory physics courses is to throw a particle up  from a certain height $h$, with initial velocity $V_{\rm{in}}$, and find the distance travelled in $x$ direction before the particle hits the ground $(y_{\rm{fin}}= 0)$. For that purpose we should first find the time of flight. Since $y_{\rm{fin}}<y_{\rm{in}}$, the quadratic equation (\ref{fall2}) gives us two solutions, one of which is negative. We usually say that the negative solution is unphysical and proceed with the positive solution for the time of flight.

In this work we take a further step to provide a deeper understanding of the problem. For that purpose we construct the group theoretical description of free fall and projectile motion.  We evaluate the group orbits regarding their relation to the physical orbits of the particles and unphysical solutions. This method has advantages. In particular the meanings of the unphysical solutions and negative time become clear.
\section{Group theoretical approach} \label{group}
A group is a set of elements with a product rule that satisfies closure, associativity, and the requirements of the existence of an identity element and an inverse for each element. This rather abstract notion of groups became relevant in theoretical physics with the development of quantum mechanics and particle physics. The gauge field theories in high energy physics are also associated with a particular gauge group. Classical problems including harmonic oscillators and Coulomb potential were also treated with the group theoretical approach \cite{group1,group2,group3,group4,group5,group6,group7,group8,group9,group10,group11,
group12}.

One of the efficient ways to analyse a dynamical system is to construct a phase space. Every possible state of a system corresponds to a unique point in the phase space. The dimension of the phase space is equal to the number of degrees of freedom for the system, and the evolution of the system is described by the path it traces through the phase space. The most common example is the phase space in Hamiltonian mechanics which is constructed by the generalized coordinates and momenta. One can also construct a phase space to describe the free fall problem. Let us consider a particle in free fall where $y$ denotes the height of the particle, and $V$ denotes its velocity  in $y$ direction at the same instant. The points $u=(y,V)$ form a two dimensional space $U \subset \mathbb{R}^2$, which is the phase space corresponding to the free fall problem. We are going to show that the kinematic equations for constant acceleration (\ref{fall1}) and (\ref{fall2}), form a one parameter group acting on this space. We define the group elements $\phi_t$ for $t \in \mathbb{R}$, by their action on $u \in U$.
\begin{equation}
\phi_t[u(t_o)]=u(t_0 + t)
\end{equation}
where $u(t_0)=( y(t_0),V(t_0) )$ and $u(t_0+t)=( y(t_0+t),V(t_0+t) )$. Explicitly one can write
\begin{eqnarray}
&& y(t_0+t)=\frac{at^2}{2} + V(t_0)t+ y(t_0) \nonumber \\
&& V(t_0+t)= V(t_0) + at \label{tzeroplust}
\end{eqnarray}
The group multiplication is defined by
\begin{equation}
\phi_{t}\circ \phi_s=\phi_{t+s}
\end{equation}
One can easily verify that closure and associativity properties are satisfied by the additive structure of the group. Trivially, the case $t=0$ corresponds to the identity element. However, it is non-trivial to show that $\phi_t$ and $\phi_{-t}$ are the inverse elements of each other, i.e.
\begin{equation}
\phi_t\phi_{(-t)}[u(t_o)]=\phi_{(-t)}\phi_t [u(t_0)]=u(t_0) \label{inverse}
\end{equation}
Note that 
\begin{eqnarray}
\phi_{(-t)}\phi_t [u(t_0)]&=& \phi_{(-t)}\left(y(t_0 + t), V(t_0 + t) \right) \nonumber \\
&=&\left( \frac{a(-t)^2}{2} + V(t_0 + t) (-t) + y(t_0 +t), V(t_0 + t)+ a(-t) \right) \nonumber \\
&=& \left( y(t_0),V(t_0) \right) 
\end{eqnarray}
where we have used the expressions for $y(t_0 + t)$ and $V(t_0 + t)$ given in (\ref{tzeroplust}). Proceeding the same way, one can also show that $\phi_t\phi_{(-t)}[u(t_0)]=u(t_0)$. Thus, the kinematic equations acting on the space $U$ generate a one parameter group $\phi_t$, such that $\phi_0$ and $\phi_{-t}$ correspond to the identity and the inverse element, respectively.

The physical meaning of $\phi_{-t}$ should be clarified. Let us consider a specific point $u(t_i)=(y(t_i),V(t_i))$. We can act on $u(t_i)$ with $\phi_t$ to find the position and the velocity in the future, $t$ seconds after $t_i$. By means of (\ref{inverse}), we can also act on $u(t_i)$ with $\phi_{-t}$ to find the position and the velocity in the past, $t$ seconds before $t_i$. The kinematic equations for constant acceleration work both to future and past. This property is essential to construct the group theoretical description.
\subsection{Generalization to projectile motion}
The group theoretical approach can be generalized to include the projectile motion. For this purpose we consider the three dimensional space $U$ with coordinates $u=(x,y,V)$, where $V$ is the velocity in $y$ direction. The group elements are defined in the same way: $\phi_t[u(t_o)]=u(t_0 + t)$. The expressions for $y(t_0+t)$ and $V(t_0+t)$ are identical with their form in (\ref{tzeroplust}). We only need to define $x(t_0 +t)$
\begin{equation}
x(t_0 +t)=x(t_0) + Ct
\end{equation}
where $C$ is the constant velocity in $x$ direction. One can easily verify that $\phi_0$ and $\phi_{-t}$ correspond to the identity and the inverse element. Then, the elements $\phi_t$ form a one parameter group acting on the phase space $U$, which describes projectile motion.

\section{Group orbit, physical orbit, unphysical solutions} \label{unphysical}
The orbit of a certain point is defined as the set of all points that can be reached by the action of the group on that point.
\begin{equation}
\mathbb{O}(u_i) \equiv \{\phi_t[u_i]; t \in \mathbb{R} \}
\end{equation}
where $\mathbb{O}$ denotes the orbit of a point. The orbit of the point $u_i$ is the set of all points that lie to the future and past of $u_i$. (including $u_i$ itself for $t=0$.) Note that $\mathbb{O}(u_i)$ is identical with  $\mathbb{O}(u_j)$ if $u_i=\phi_t [u_j]$, where $t$ can be positive or negative. On the other hand if the point $u_a$ cannot be reached by the action of the group on $u_i$, the orbits of $u_a$ and $u_i$ do not have a common element. In other words two group orbits are either identical or disjoint. Every point in the phase space lies in a unique orbit. The phase space is constructed by the union of all orbits. It is called a disjoint union, since two different orbits do not have a common element.

In physical problems of free fall or projectile motion, by specifying a single point in the phase space we restrict ourselves to the orbit of that point. The group orbit includes all the points from $t=-\infty$ to $t=\infty$. On the other hand the physical problem is defined in an interval $t=[t_{\rm{in}},t_{\rm{fin}}]$. (Usually we let $t_{\rm{in}}=0$)  The physical orbit of a particle in free fall or projectile motion corresponds to a connected subset of the group orbit. A specific problem is identified by a single point in the phase space which determines the group orbit, and an interval $[t_{\rm{in}},t_{\rm{fin}}]$. The points $u_t$ in the same orbit are unphysical solutions, if $t<t_{\rm{in}}$ or $t>t_{\rm{fin}}$. 

One tends to identify the points with a negative time parameter as unphysical. In fact these two are independent concepts. The confusion stems from the fact that we usually assign $t=0$ to our initial point. Then the points that lie to its past acquire a negative time parameter; at the same time they  become unphysical as they are not included in the subset corresponding to our physical problem, i.e. $t<t_{\rm{in}}$. Without loss of generality, we can also assign $t_{\rm{fin}}=0$. Then the points in the interval $t=[t_{\rm{in}},t_{\rm{fin}})$ will  have a negative time parameter though they are physical. The time parameter of a point becomes positive or negative depending on whether it lies to the future or past of the point that we assign $t=0$. As we have mentioned we can assign $t=0$ to any point in the phase space.

Let us clarify our arguments with a simple example. Consider, a particle that is released  from a height $h$ and allowed to fall freely. In this example we restrict ourselves to the orbit of the point $u_0 =(h,0)$. As usual we assign $t=0$ to this point, i.e. $u(t=0)=(h,0)$. For $t<0$ the points in the group orbit $u_t=\phi_t [u_0]$ are unphysical since $t<t_{\rm{in}}$. We find that $\phi_{t_{\rm{fin}}}[u_0]=(0,-\sqrt{2hg})$, where $t_{\rm{fin}}={\sqrt{(2h)/g}}$. Our particle hits the ground with a velocity $V_{\rm{fin}}=-\sqrt{2hg}$ at $t_{\rm{fin}}=\sqrt{(2h)/g}$. The points $u_t=\phi_t[u_0]$ are also unphysical for $t>\sqrt{(2h)/g}$. We can act on any point with $\phi_t$ to find the position and velocity in its past or future $\phi_t[u(t_a)]=u(t_a+t)$. Here $t$ can be positive or negative. The solution is unphysical if $t_a+t<0$ or $t_a+t>\sqrt{(2h)/g}$. Our free fall problem is identified by the point $u_0=(h,0)$ and the interval $[t_{\rm{in}}=0,t_{\rm{fin}}=\sqrt{(2h)/g}]$. Equivalently we can use any other point in the same orbit to identify our problem.

\section{Conclusions}
In this work we formulated the group theoretical description of free fall and projectile motion. For that purpose we constructed a phase space with coordinates $(y,V)$ and showed that the kinematic equations acting on this space form a one parameter group which describes free fall near the surface of the Earth or any constant acceleration problem in general. The group elements $\phi_t$ act on the points $u(t_a)$ to give $u(t_a +t)$, where $t$ can be positive or negative. We generalized this argument to include projectile motion, with a three dimensional phase space $(x,y,V)$. We argued that every free fall or projectile motion problem can be identified by a single point in the phase space and an interval $[t_{\rm{in}},t_{\rm{fin}}]$. The point in the phase space determines the group orbit, since every point lies in a unique orbit. The group orbit includes all the points from $t=-\infty$ to $t=\infty$, while the physical orbit is restricted to the interval $[t_{\rm{in}},t_{\rm{fin}}]$. Usually we assign $t=0$ to our initial point. In fact we can assign $t=0$ to any point in the phase space without loss of generality. In this approach the meaning of the unphysical solutions also becomes clear. Unphysical solutions simply correspond to the points in the specified orbit, which are not included in the interval $[t_{\rm{in}},t_{\rm{fin}}]$ for the relevant problem.

The group theoretical formulation derived in this work does not apply to a more general problem involving a time dependent acceleration. Let us consider the simplest case $a=a_0 + c_1t$. The expressions for $y$ and $V$ take the form:
\begin{eqnarray}
&& y(t_0+t)=\frac{c_1 t^3}{6}+ \frac{a_0 t^2}{2} + V(t_0)t+ y(t_0) \nonumber \\
&& V(t_0+t)=\frac{c_1 t^2}{2}+ a_0t+ V(t_0) 
\end{eqnarray}
In this case we have
\begin{eqnarray}
&&\phi_{(-t)}\phi_t [u(t_0)]= \phi_{(-t)}\left(y(t_0 + t), V(t_0 + t) \right) \nonumber \\
&=&\left(\frac{c_1 (-t)^3}{6}+ \frac{a_0(-t)^2}{2} + V(t_0 + t) (-t) + y(t_0 +t), \frac{c_1 (-t)^2}{2} + a(-t) + V(t_0 + t) \right) \nonumber \\
&=&\left( \frac{-c_1 t^3}{2} + y(t_0), c_1 t^2 + V(t_0) \right) \neq \left( y(t_0),V(t_0) \right) 
\end{eqnarray}
Since $\phi_{-t}\phi_t [u(t_0)]\neq u(t_0)$, $\phi_{-t}$ cannot be identified as the inverse element of $\phi_t$. Physically, we cannot say that $\phi_{-t}$ generates translations to the past. Constant acceleration is a special case where $\phi_{-t}$ generates translations to the past so that one can define the inverse element. Obviously, the formulation applies to  the case $a=0$, namely the constant velocity problems.

This method improves our understanding of the constant acceleration problem with its global approach. It also provides a pedagogical example which will be useful in teaching the abstract concepts of group theory. It is a simple demonstration of how to use abstract mathematical concepts to describe the physics of daily life. In this sense, it is especially beneficial for students who want to pursue a career in theoretical physics.

%

\end{document}